\documentclass[prl,amsmath,amssymb,aps,twocolumn]{revtex4-2}

\usepackage{amsfonts}
\usepackage{amsmath}
\usepackage{amsthm}
\usepackage{graphics}
\usepackage{graphicx}
\usepackage{subfigure}
\usepackage{amssymb}
\usepackage{graphics}
\usepackage{graphicx}
\usepackage{epstopdf}
\usepackage{color}
\usepackage{subfigure}
\usepackage{pgfplots,tikz}
\usepackage{url}
\usepackage{soul}
\usepackage{enumitem}
\usepackage{lipsum}
\usepackage{hyperref}
\usepackage{cleveref}
\usepackage{comment}
\usepackage{enumitem} % for list customization

\usepackage{hyperref}
\hypersetup{
    colorlinks=true,
    linkcolor=blue,
    filecolor=magenta,      
    urlcolor=blue,
    }

\usepackage{titlesec}

\definecolor{astral}{RGB}{46,116,181}

\titleformat{\section}
{\color{blue}\normalfont\Large\bfseries\center\uppercase }
{\color{blue}\thesection}{1em}{}

\titleformat{\subsection}
{\normalfont\bfseries\center\uppercase }
{\thesection}{1em}{}

\titleformat{\subsubsection}
{\normalfont\bfseries\center }
{\thesection}{1em}{}

\def\k{\textbf{k}}

\newcommand{\MyItem}[1]
{\vskip 0.25cm
\emph{\textbf{- {#1}}}
\vskip 0.25cm
}

\newcommand{\WavKinS}{\texttt{WavKinS.jl }}
\newcommand{\WavKinSns}{\texttt{WavKinS.jl}} %WavKinS.jl without space

\begin{document}

\title{\color{blue} \WavKinSns : an efficient and modular Julia software for solving wave kinetic equations}

\author{Giorgio Krstulovic}
\affiliation{Universit\'{e} C\^{o}te d'Azur, Observatoire de la C\^{o}te d'Azur, CNRS, Laboratoire Lagrange, Boulevard de l'Observatoire CS 34229 -- F 06304 Nice Cedex 4, France}
\author{Vincent Labarre}
\affiliation{LadHyX (CNRS UMR 7646), École Polytechnique, 91128 Palaiseau, France \\
Universit\'{e} C\^{o}te d'Azur, Observatoire de la C\^{o}te d'Azur, CNRS, Laboratoire Lagrange, Boulevard de l'Observatoire CS 34229 -- F 06304 Nice Cedex 4, France}

\begin{abstract}
    This article describes the Wave Kinetic Solver \WavKinSns, a software developed in the Julia language for solving different types of wave kinetic equations in an efficient and modular manner. \WavKinS already solves the wave kinetic equation describing the interaction of waves in several physical systems, such as acoustic waves, Bose-Einstein condensates, internal gravity waves and many others. Thanks to the structures and routines already implemented in \WavKinSns, we expect that developing a new solver for a specific physical system to be straightforward. \WavKinS is an Open Source project distributed with EUPL-1.2 license agreement.\\
    \center{\url{https://github.com/gkrstulovic/WavKinS.jl}\\
    \url{https://gkrstulovic.github.io/WavKinS.jl/}}
\end{abstract}

\maketitle

\begin{figure}[h!]
    \includegraphics[width=0.45\columnwidth]{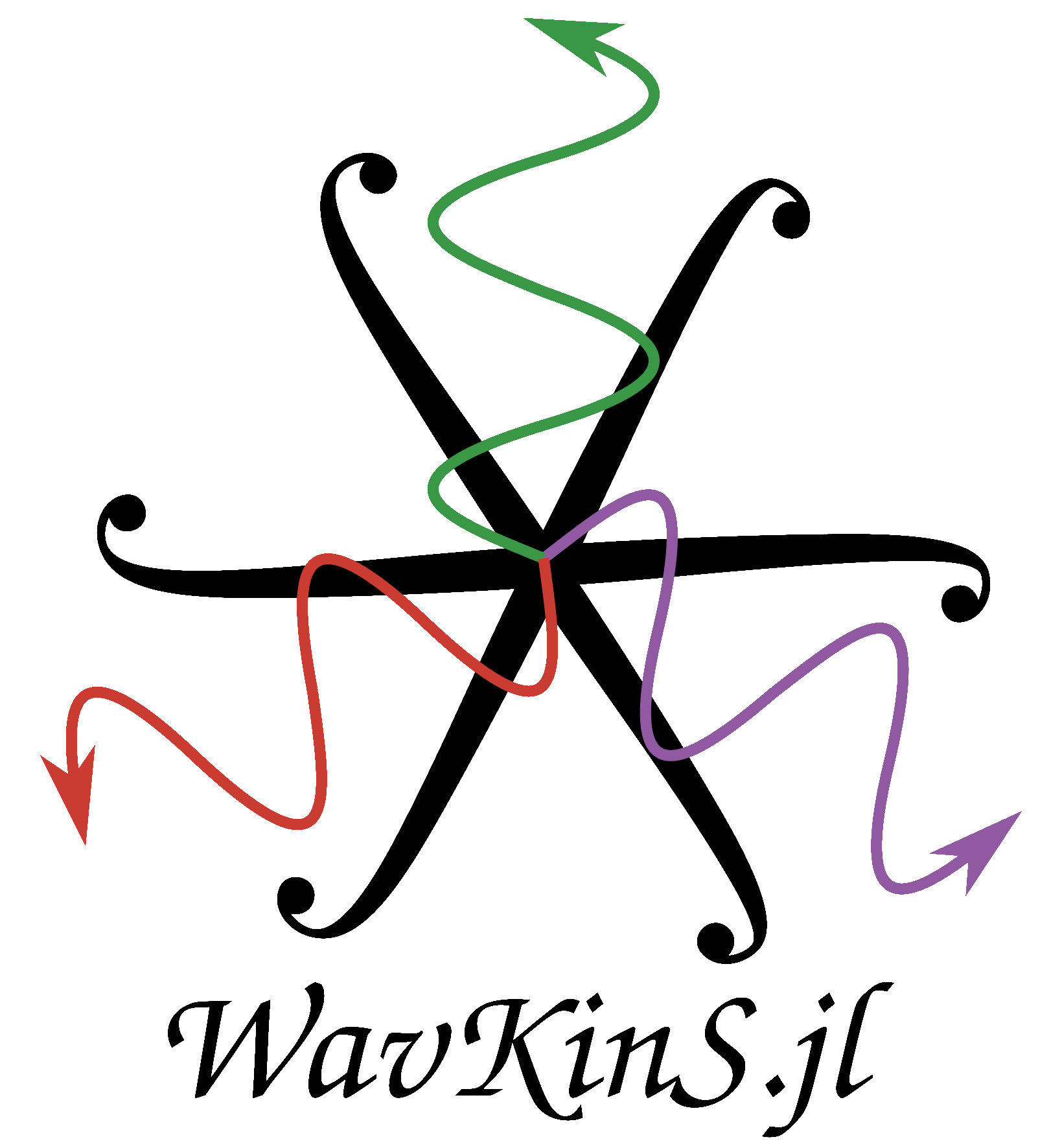} 
    \caption{\WavKinS logo.}
    \label{fig:logo}
\end{figure}
\vspace{-2cm}
\section{Overview}

\subsection{Introduction}

\subsubsection{General description}

Wave kinetic equations arrive in many physical systems, such as plasmas, geophysical fluids, elastic plates, turbulent Bose-Einstein condensates, and many others \cite{Galtier_WeakInertialwaveTurbulence_2003,Lvov_WeakTurbulenceKelvin_2010,dyachenko1992optical,zhuDirectInverseCascades2023}. When the nonlinearity is small, the theory of Weak Wave Turbulence (WWT) provides a close equation describing the evolution of the wave spectrum, which relates to the wave amplitude or energy at a given scale \cite{ZLF,nazarenko2011wave}. 

This article does not aim to provide an introduction to WWT, we provide however the physical context where wave kinetic equations are found. Let's start by considering a general non-linear wave system, where the amplitudes of the Fourier modes $a_\k$, having a linear dispersion relation $\omega_\k$ are described by the following Hamiltonian
\begin{eqnarray}
 \nonumber   H&=&\sum_{\k} \omega_\k |a_\k|^2 + \sum_{\bf 123} V_{\bf 123} ( a_{\bf 1}^* a_{\bf 2}a_{\bf 3} +c.c.)\delta^{\bf 1}_{\bf 23}\\
    &&+ \sum_{\bf 1234} T^{\bf 12}_{\bf 34}  a_{\bf 1} a_{\bf 2}a_{\bf 3}^*a_{\bf 4}^*\delta^{\bf 12}_{\bf 34} \label{eq:Hamiltonian}
\end{eqnarray}
Here, we consider waves in a $d$-dimensional periodic box of size $L$, such that wavevectors $\k\in \frac{2\pi}{L}\mathbb{Z}$. We denote $a_{\bf i}=a_{\bf k_i}$ and use the same convention for summation indexes. $a_\k^*$ denotes the complex conjugate of $a_\k$. The tensors $V_{\bf 123}$ and $T^{\bf 12}_{\bf 34}$ are the interaction matrices that contain all the physics of the system, which for simplicity we consider here fully symmetric. Finally, $\delta^{\bf 1}_{\bf 23}$ and $\delta^{\bf 12}_{\bf 34}$ are $\delta$-Kronecker such that they are non-zero and equal to one only if $\k_1=\k_2+\k_3$ and $\k_1+\k_2=\k_3+\k_4$, respectively. Note that the wave dynamics is simply given by $i\dot{a_\k}=\frac{\partial H}{\partial a_\k^*}$.

The WWT provides a statistical description for the waveaction spectrum $n_\k=\frac{L^d}{(2\pi)^d}\langle |a_\k|^2 \rangle$ in the limit of infinite box size and small nonlinearity (taken in that order). Rigorously, $\langle \,\,\rangle$ denotes an average over initial conditions, although it is often considered as an average over different sources of fluctuations in the system. The statistical description of waves in the above mentioned limit is known as the kinetic description of weakly nonlinear waves. Note that if $V_{\bf 123}\neq0$, then the quartic terms in the Hamiltonian can be neglected in this limit.
In the end, the WWT furnishes a closed equation for the waveaction spectrum called the wave kinetic equation (WKE)
\begin{equation}
    \frac{\mathrm{d}n_{\bf k}}{\mathrm{d}t} = St_{\bf k},
\end{equation} 
where $St_{\bf k}$ is the collisional integral. The collisional integral is a nonlinear integral functional of $n_{\bf k}$ defined over a manifold, which will be discussed later. 
Finally note that in the Hamiltonian \eqref{eq:Hamiltonian}, one could have considered more general terms, such as $U_{\bf 123}( a_{\bf 1}a_{\bf 2}a_{\bf 3} +c.c.)$, however the WWT shows that in the kinetic limit, only those present in \eqref{eq:Hamiltonian} survive and contribute to the dynamics of $n_\k$. Moreover, the Hamiltonian structure of the nonlinear wave system considerably simplifies the analytical computations, but it is not necessarily required in the derivation of the WKE.

The WKE can be classified by the number of waves interacting, which results in different degrees of nonlinearity and integrations to be performed. In the current version of \WavKinSns, we consider two types of systems that can be generically written as follows.

%\subsubsection{$3$-waves canonical systems}
\MyItem{$3$-wave canonical systems:}

In this case, the leading nonlinearity in \eqref{eq:Hamiltonian} is cubic, and the collisional integral reads
\begin{equation}
St_{\bf k}=\int (\mathcal{R}^{\bf k}_{12}-\mathcal{R}^{1}_{2 {\bf k}}-\mathcal{R}^{2}_{{\bf k}1})\mathrm{d}{\bf k}_1\mathrm{d}{\bf k}_2 \label{eq:3W_KE}
\end{equation}
with 
\begin{eqnarray}
\nonumber\mathcal{R}^{\bf k}_{12}=&&2\pi|V_{{\bf k}12}|^2(n_1n_2-n_1n_{\bf k}-n_{\bf k}n_2)\\
 &&\times\delta({{\bf k}-{\bf k}_1-{\bf k}_2})\delta(\omega_{\bf k}- \omega_1- \omega_2). 
\end{eqnarray}
The integrals in the collisional integral are performed over the so-called resonant manifold defined by two $\delta$-Dirac, 
\begin{equation}\label{eq:3W_Res_manifold}
    {\bf k}={\bf k}_1+{\bf k}_2,\quad \omega_{\bf k}= \omega_{\bf 1}+ \omega_{\bf 2},
\end{equation}
and permutation of indexes.

The $3$-waves WKE conserves the total energy of the system given by
\begin{equation}
    E=\int\omega_{\bf k} n_{\bf k} d{\bf k}.\label{eq:energy}
\end{equation}
Depending on the physical system, other invariants might exist.

Note that if wavevectors are in dimension $d$, then there are $2d - d - 1=d-1$ integrals to be performed for each value of a external wavevector ${\bf k}$. Therefore, if we use $M$ points to discretise each coordinate of the wavevector space, then the numerical cost will be generically of the order of $M^{2d -1}$ operations. However, by assuming full isotropy or isotropy in a plane one can drastically reduce the numerical cost, making calculations feasible.

%\subsubsection{$4$-waves canonical systems}
\MyItem{$4$-wave canonical systems:}

In these systems, the cubic term in \eqref{eq:Hamiltonian} is absent or the resonant manifold \eqref{eq:3W_Res_manifold} is empty. In this case, the collisional integral is
\begin{eqnarray}
\nonumber St_{\bf k}=&&\int |T^{{\bf k 1}}_{\bf 2 3}|^2\delta({{\bf k}+{\bf k}_1-{\bf k}_2}-{\bf k}_3)\delta(\omega^{{\bf k 1}}_{\bf 2 3}) n_{\bf k}n_{\bf 1}n_{\bf 2}n_{\bf 3}\\
&&\times\left( \frac{1}{n_{\bf k}} + \frac{1}{n_{\bf 1}} -\frac{1}{n_{\bf 2}} -\frac{1}{n_{\bf 3}} \right)\mathrm{d}{\bf k}_1\mathrm{d}{\bf k}_2\mathrm{d}{\bf k}_2\label{eq:4W_KE}
\end{eqnarray}
with $\omega^{{\bf k 1}}_{\bf 2 3}=\omega_{\bf k}+ \omega_{\bf 1}- \omega_{\bf 2}-\omega_{\bf 3}$. The resonant manifold is in this case
\begin{equation}\label{eq:4W_Res_manifold}
    {{\bf k}+{\bf k}_1-{\bf k}_2}-{\bf k}_3 = 0,\quad \omega_{\bf k}+ \omega_{\bf 1}- \omega_{\bf 2}-\omega_{\bf 3}=0
\end{equation}

Besides the energy \eqref{eq:energy}, $4$-waves WKEs also conserves the total waveaction of the system
\begin{equation}
    N=\int n_{\bf k} d{\bf k}.\label{eq:waveaction}
\end{equation}

For $4$-wave systems there are $3d-d-1=2d-1$ integrals to be performed, which implies in general a numerical cost of the order of $M^{3d-1}$ operations per time step.

\subsubsection{The Wave Kinetic Solver \WavKinSns}

The Wave Kinetic Solver (\WavKinSns) is a software developed in Julia language to solve WKEs of the kind \eqref{eq:3W_KE} and \eqref{eq:4W_KE}. More precisely, it solves the following type of equations
\begin{equation}
    \frac{\mathrm{d}n_{\bf k}}{\mathrm{d}t} = St_{\bf k} + f_{\bf k} - d_{\bf k} n_{\bf k},
\end{equation}
where $f_{\bf k}$ is a forcing term and $d_{\bf k}\ge0$ a dissipation coefficient. Their definitions depend on the physical problem. 

\WavKinS provides all the structures and functions needed to solve several types of wave kinetic equations. The code runs well on personal computers and shared-memory clusters. \WavKinS is conceived in such a way that implementing a new physical system should be straightforward, provided that some basic mathematical knowledge of the physical system has been previously achieved (e.g., parametrisation of resonant manifolds, study of possible singularities of the integrands, etc.). \WavKinS is available in \url{https://github.com/gkrstulovic/WavKinS.jl} and its documentation here \url{https://gkrstulovic.github.io/WavKinS.jl/}. \WavKinS is an Open Source project distributed with EUPL-1.2 license agreement.

In the following, we provide the list of currently implemented physical systems in \WavKinSns. The specific details about their implementation can be found in the documentation.

\MyItem{Acoustic waves}

\WavKinS solves the isotropic wave kinetic equation for acoustic waves in two and three dimensions. The dispersion relation is $\omega_k=c|k|$ with $c$ the speed of sound. The interaction matrix is $V_{123}=V_0 \sqrt{k_1k_2k_3}$. The acoustic WKE is theoretically derived from a weakly dispersive system with dispersion relation $\omega_k=ck(1+a^2k^2)$ in the limit $a\to0$ \cite{ZLF,griffinEnergySpectrumTwoDimensional2022,zhu2024turbulence}. The collisional integral of the WKE is
\begin{equation}
    St_{k}=\frac{C_d}{a^{3-d}k^{d-1}}\int \left(\mathcal{N}^k_{12}-\mathcal{N}^1_{2k}-\mathcal{N}^2_{k1}\right)(k k_1k_2)^{d-2}\mathrm{d}k_1\mathrm{d}k_2,
\end{equation}
with $\mathcal{N}^k_{12}=|V_{k12}|^2(n_1n_2-n_1n_k-n_kn_2)\delta(\omega_k-\omega_1-\omega_2)$, and where $d=2$ or $3$ is the dimension. The prefactor is $C_2=2\pi/\sqrt{6}$ and $C_3=4\pi^2$. 

Because isotropy is assumed, the numerical cost is of the order of $M^2$ operations per time step.

\MyItem{Three-dimensional Bogoliubov waves}

Bogoliubov waves are elementary excitations of Bose-Einstein condensates obtained from the linearised Nonlinear Schrodinger equation around a flat condensate state. The $3D$ isotropic WKE collisional integral is
\begin{equation}
    St_{k}=\frac{4\pi}{k^2}\int \left(\mathcal{N}^k_{12}-\mathcal{N}^1_{2k}-\mathcal{N}^2_{k1}\right)k k_1k_2\mathrm{d}k_1\mathrm{d}k_2,
\end{equation}
with $\mathcal{N}^k_{12}$ as for acoustic waves, but with the Bogoliubov dispersion relation $\omega_k=ck\sqrt{1+\frac{1}{2}\xi^2k^2}$. For this system, the interaction matrix is 
\begin{equation}
V_{12}^k = V_0\sqrt{kk_1k_2} \left[ \frac{1}{2\sqrt{\eta_k \eta_1 \eta_2}} + \frac{\sqrt{\eta_k \eta_1 \eta_2}}{6 k k_1 k_2} \left( \frac{k^3}{\eta_k} - \frac{k^3}{\eta_1} - \frac{k^3}{\eta_2} \right) \right],
\end{equation}
with $\eta_k = \sqrt{1+(k\xi)^2/2}$. Note that Bogoliubov WKE reduces to the acoustic WKE in the limit $\xi\to0$. It has the same numerical cost than the acoustic WKE.

\MyItem{MMT waves}

The Majda-McLaughlin-Tabak (MMT)  model was introduced by the authors as a simple and numerically affordable model to test the theory of wave turbulence \cite{majdaOnedimensionalModelDispersive1997}. It consists on a one-dimensional partial differential equation where the dispersion relation and the homogeneity degree of the non-linearity are tunable parameters. This model has triggered enormous research and opened the way to the study of solitonic turbulence. It is still the matter of recent academic research, both from mathematicians and theoretical physicists. For a complete review see \cite{zakharovOnedimensionalWaveTurbulence2004}.

\WavKinS solves the 1D-MMT WKE \eqref{eq:4W_KE} with 
\begin{eqnarray}
    \omega_k=k^{1/2},\quad {\rm and}\quad T^{12}_{34}=|k_1k_2k_3k_4|^{\beta/4}.
\end{eqnarray}
The full parametrisation of the resonant manifold is provided in the documentation. The numerical cost of the 1D-MMT WKE is of the order of $M^2$ operations per time step.

\MyItem{Nonlinear Schrödinger equation}

The nonlinear Schrodinger (NLS) equation is one of the most important model in physics. Its applications vary from nonlinear optics, water waves to Bose-Einstein condensates. The equation is also known as the Gross-Pitaevskii equation, a fundamental equation describing superfluids. For a discussion on the wave kinetic description of the NLS equation, see \cite{dyachenko1992optical,nazarenko2011wave,zhuDirectInverseCascades2023}.

\WavKinS solves the isotropic 3D-NLS WKE, which corresponds to taking $\omega_k=k^2$ and $T^{12}_{34}=1$ in Eq.~\eqref{eq:4W_KE}. Assuming isotropy, the 3D-NLS WKE collisional integral reduces to
\begin{eqnarray}
 \nonumber St_k&=&   \frac{4\pi^3}{k} \int \mathcal{S}^{k1}_{23}\delta(\omega^{k1}_{23})\, n_kn_1n_2n_3\\
 &&\times\left(\frac{1}{n_k}+\frac{1}{n_1}-\frac{1}{n_2}-\frac{1}{n_3}   \right)d\omega_1d\omega_2d\omega_3,
\end{eqnarray}
with $\mathcal{S}^{k1}_{23}=\min{[k,\sqrt{\omega_1},\sqrt{\omega_2},\sqrt{\omega_3}]}$.

The numerical cost of the isotropic NLS WKE is of the order of $M^3$ operations per time step.

\MyItem{Petviashvilli waves}
The Petviashvilli equation is an important model describing geophysical flows and fusion plasmas. In particular, it models drift waves in plasmas and Rossby waves in geophysical fluids \cite{nazarenko2011wave}. 

\WavKinS solves the Petviashvilli WKE for $n_{\bf k}$ and ${\bf k}=(k_x,k_y)$, with $k_x>0$ in the limit $k_x\ll |k_y|$, which corresponds to \eqref{eq:3W_KE} with
\begin{equation}
    \omega_{\bf k}=k_xk_y^2,\quad {\rm and}\quad V_{{\bf 123}}=\frac{1}{2}\sqrt{k_{1x}k_{2x}k_{3x}}.
\end{equation}
The parametrisation of the resonant manifold is provided in the documentation.
In addition to energy, the Petviashvilli WKE conserves the potential vorticity $\Omega$ and the zonostrophy $\Phi$ defined as
\begin{equation}
\Omega = \int_{k_x>0} k_x n_{\bf k} d{\bf k},\quad \Phi= \int_{k_x>0} \frac{k_x^3}{k_y^2} n_{\bf k} d{\bf k}.
\end{equation}

The numerical cost of the WKE is of the order of $M^3$ operations per time step.

\MyItem{Stratified waves}

Internal Gravity Waves (IGW) propagate in stratified flows like the oceans and atmospheres. They are therefore important for geophysical applications (see \cite{lvovResonantNearResonantInternal2012a} for a review). \WavKinS solves the IGW WKE in the strong anisotropy (or hydrostatic) limit, where we consider that the horizontal wave vector $k_h$ is small compared to the vertical wave vector $k_z$.   

The WKE for internal waves in the hydrostatic limit is given by equation \eqref{eq:3W_KE} with
\begin{eqnarray}
    V_{k12} = \sqrt{\frac{k_h k_{1h} k_{2h}}{32}} \left( \frac{{\bf k}_h \cdot {\bf k}_{1h}}{k_h k_{1h}} \sqrt{\left| \frac{k_{2z}}{k_z k_{1z}} \right|} +\,permutations\,\right),
\end{eqnarray}
and $\omega_k=N k_h/|k_z|$ with $N$ the buoyancy frequency \cite{lvovHamiltonianFormalismGarrettMunk2001,Labarre_Lanchon_Cortet_Krstulovic_Nazarenko_2024}%\citelvovHamiltonianFormalismGarrettMunk2001,labarreKineticsInternalGravity2024}.

\WavKinS assumes isotropy on the plane perpendicular to the vertical direction, which reduces the numerical cost of the WKE to roughly $M^4$ operations per time step. The numerical integration is therefore costly, but feasible in shared memory clusters \cite{labarre2024wave}. The equations solved by \WavKinS are explicitly given in the documentation.

\MyItem{Smoluchowski kinetic equation}

The Smoluchowski equation is a standard model of statistical physics. It describes coagulation processes such as droplets in clouds, blood clots, meteorite formation from stellar dust, and so on.  Although it is not a wave kinetic equation, it shares a similar structure with the WKE, so all the routines implemented in \WavKinS can be used. 

For the Smoluchowski equation, $k$ denotes the particle mass and $n_{k}$ is the particle mass distribution. The collision integral is 
\begin{eqnarray}
\nonumber    St_{k} =&& \frac{1}{2} \int\limits_0^k ~ K(k-p,p) ~ n(k-p,t) ~ n(p,t) ~ \mathrm{d}p \\
    &&-\int\limits_0^\infty ~ K(k,p) ~ n(k,t) ~ n(p,t) ~ \mathrm{d}p.
\end{eqnarray}
Some kernels $K$, quantifying the interaction between particles, are already implemented in \WavKinSns, but the user can easily implement new ones and pass them as arguments.

Finally, the numerical cost of the Smoluchowski equation is of the order of $M^2$ operations per time step.

\subsection{Implementation and architecture}

\subsubsection{Numerical schemes}

To simulate a WKE, one must first compute the collisional integral $St_{\bf k}$, which results in computing several integrals over the resonant manifold that is assumed to be known analytically. In order to perform such integrals, one also needs to interpolate the waveaction spectrum $n_{\bf k}$ on intermesh values. Once $St_{\bf k}$ is known, the time evolution of $n_{\bf k}$ is obtained using a given time-stepping scheme.

In the following, we describe the basic methods used by \WavKinSns. For the sake of simplicity, we reduce the discussion to systems where $n_{\bf k}$ is a function of a single variable $k$ ($1d$ or isotropic spectra). The systems depending on two variables are straightforward generalisations and they are discussed in the documentation.

\MyItem{Grids and basic fields}

The basic structure of \WavKinS is the \texttt{wave\_spectrum} Julia structure, which contains the waveaction spectrum $n_k$ and the grid of wavevectors. The wavevector space is discretised using $M$ nodes in a logarithmic grid:
\begin{equation}
    k_i = k_{\rm min} \lambda^{i-1},\quad i=1,\ldots,M.
\end{equation}
The structure can be constructed in different ways specified in the documentation. 

\MyItem{Integration}

For integration, the codes uses by default a standard trapezoidal rule on the log-lattice. For instance, the integral over two mesh points is treated as
\begin{eqnarray*}
    \int_{k_{i}}^{k_{i+1}} F(k) \mathrm{d}k&=& \log{\lambda}\int_{i}^{i+1} F(k(x))k(x) \mathrm{d}x \\
    &\approx& \frac{\log{\lambda}}{2} (F_{i+1} k_{i+1}+F_{i}k_{i}),
\end{eqnarray*}
where $F_i=F(k_i)$ and we have performed the change of variable $k=k_{\rm min} \lambda^x$. For integrals on the interval $[0,k_1]$, the code uses directly a  standard trapezoidal rule, where the value of $F$ at $k=0$ is obtained by linear extrapolation. In the tests, it is shown an excellent convergence of the method. Other integration schemes are also available.

\MyItem{Interpolation}

The default interpolation method is linear. It has the advantage of preserving positivity, a key property to ensure stability of the integration of the WKEs. Several other schemes are available, but they are less robust. 

\begin{figure}[h!]
    \includegraphics[width=0.99\columnwidth]{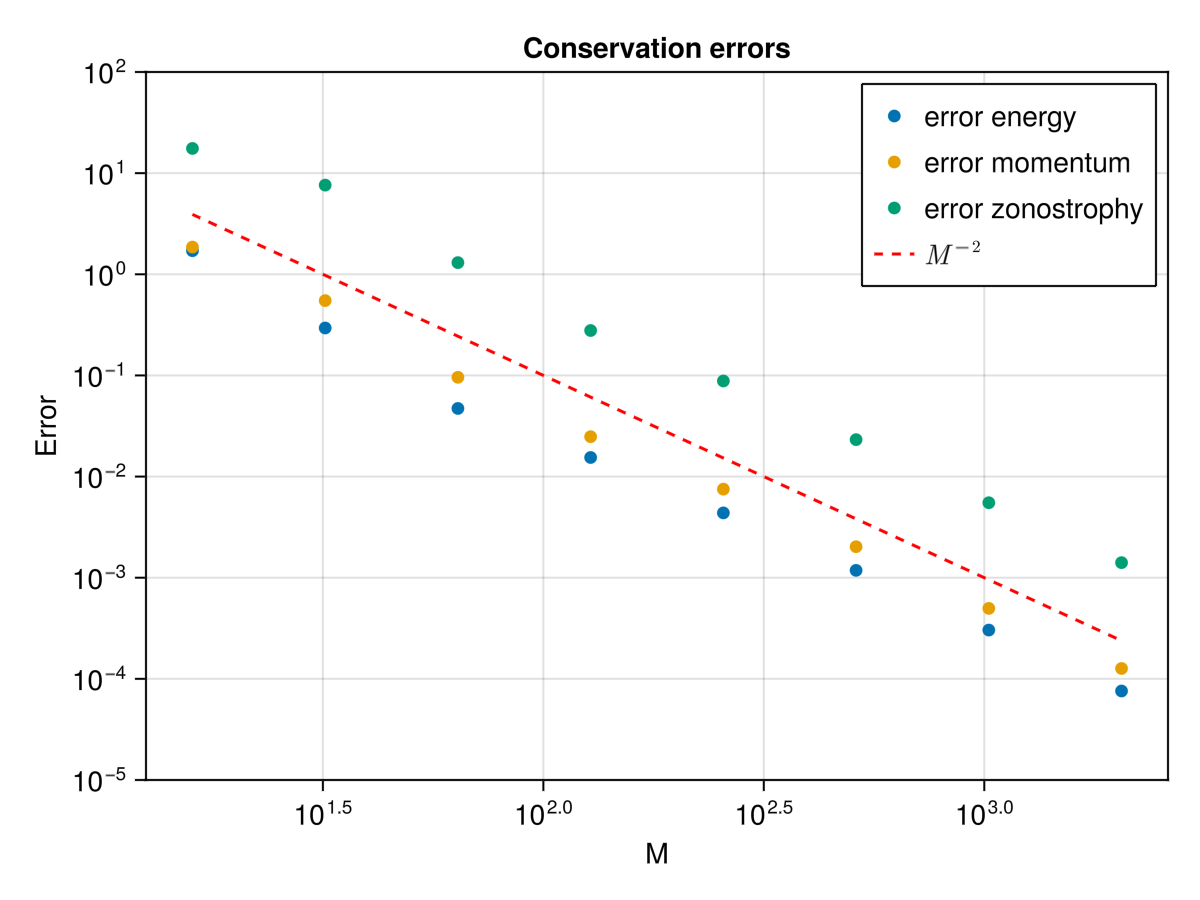} % Adjust width or height as needed
    \caption{Conservation errors of energy, potential vorticity and zonostrophy for the Petviashvilli WKE. The error is computed as $Error=(Q(t)-Q(0))/Q(0)$, where $Q(t)$ is one of the invariants. The figure is taken from \WavKinS documentation.}
    \label{fig:conservation}
\end{figure}

The errors in the computation of the WKE mainly come from interpolations, for instance, when using linear interpolation, the error in the conservation of invariants decrease as $M^{-2}$, where $M$ is the number of mesh points per dimension. As a matter of example, in Fig.~\ref{fig:conservation} we show the conservation errors for the Petviashvilli WKE invariants.   
For each physical system, the documentation contains a similar study of the invariant conservation. The respective scripts are also provided in the code.

\MyItem{Temporal schemes}

For all physical systems, several standard temporal schemes are available in \WavKinS (e.g. Euler, RK2, RK4, etc). Some solvers, like Petviashvilli waves, also offers more elaborated splitting methods for time stepping. The codes also includes an adaptive time-stepping method which allows for saving computational time.

\subsubsection{Running the code}

\begin{figure}[h!]
    \includegraphics[width=0.8\columnwidth]{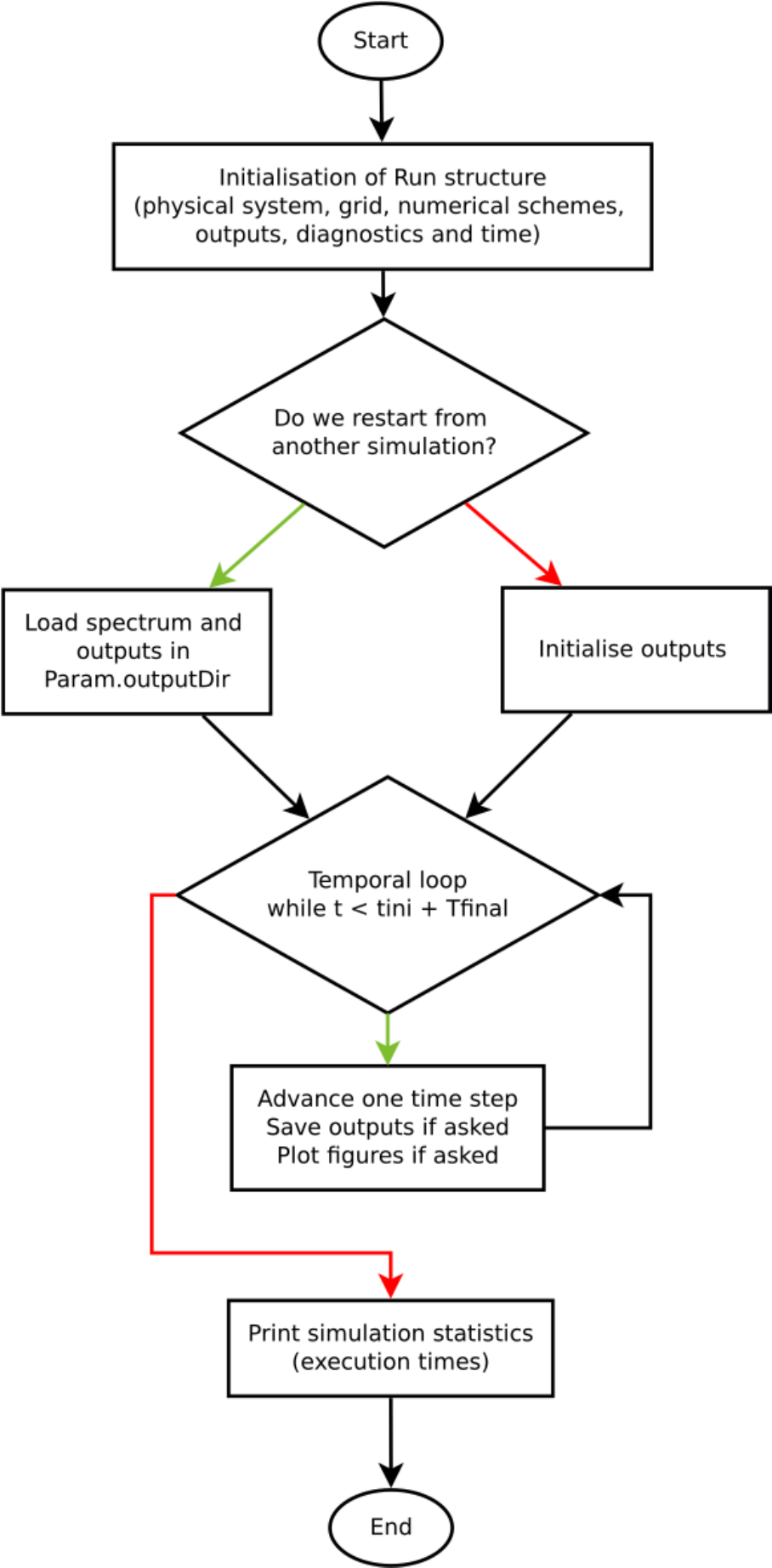} % Adjust width or height as needed
    \caption{Flowchart of the \WavKinSns}
    \label{fig:flowchart}
\end{figure}

The code contains basic Julia scripts to run simulations of the different physical systems. Once the grid are defined, the user creates a Julia structure for the physical system to be simulated. This structure contains all the necessary information, functions and fields to run the simulation. 
The flowchart of the \WavKinS is displayed in Fig.~\ref{fig:flowchart}.

\begin{figure}
    \centering
    \includegraphics[width=0.95\columnwidth]{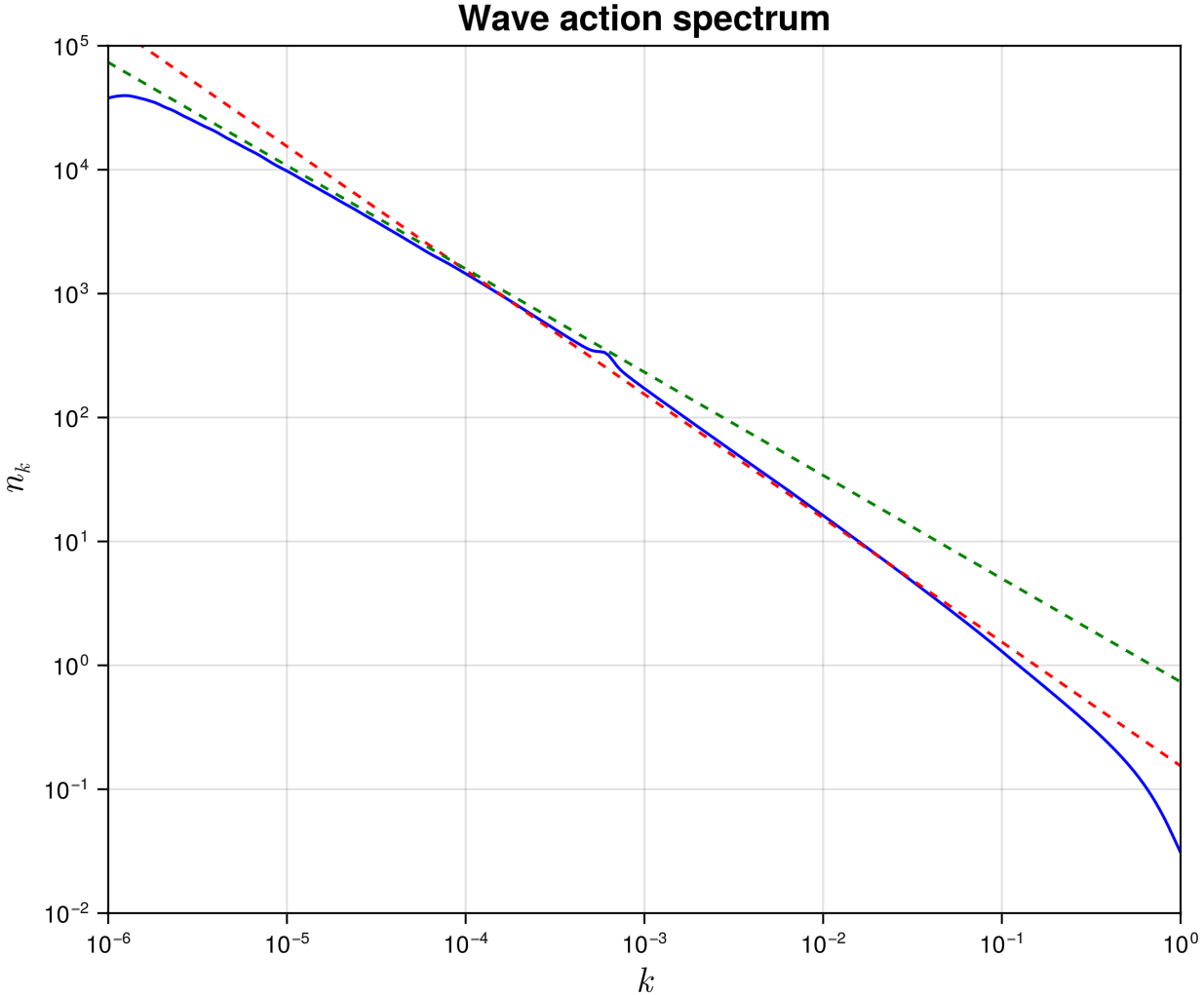}
    \includegraphics[width=0.95\columnwidth]{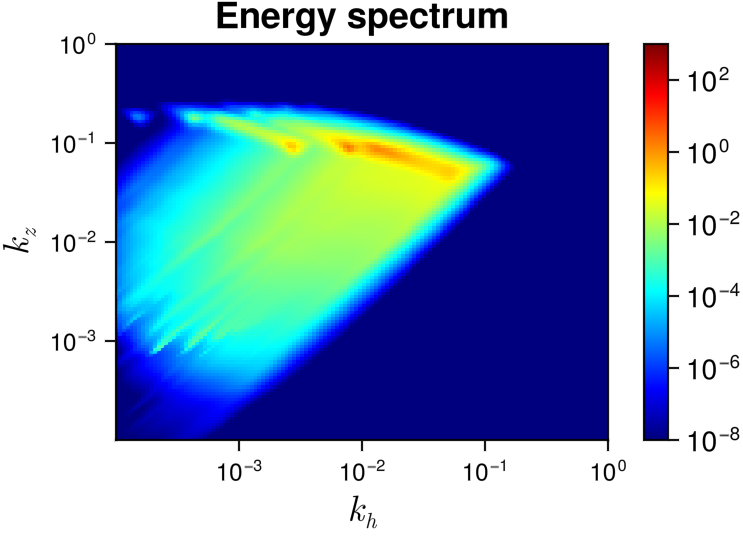}
    \caption{\textbf{Top:} Final state of the MMT WKE with forcing and dissipation. The system is forced in a narrow zone around $k_f=8\times10^{-4}$, and dissipation is added at both, small and large $k$. The two dashed line are the theoretical predictions of the WWT (see documentation). \textbf{Bottom:} Final state of the Petviashvilli WKE with forcing and dissipation. The forcing is localised around $(10^{-1},10^{-1})$. Figures reproduced form \WavKinS documentation.}
    \label{fig:example_WKE_simulations}
\end{figure}

For each physical system, \WavKinS contains running scripts showing a simple temporal evolution. For instance, Fig.~\ref{fig:example_WKE_simulations} shows the final states obtained after running the MMT and Petviashvilli WKE, with forcing and dissipation.

\subsubsection{Software architecture}

The code is written in Julia language, and organised in the standard way of Julia projects. More precisely, it is structured in the following manner:   
\begin{description}[leftmargin=.3cm]
    \item[src]Source files of the code, organised as follows:
    \begin{description}[leftmargin=.3cm]
        \item [grid]: Structures and functions for the grids.
        \item [integration] Tools used for integrations, mainly for log grids. 
        \item [interpolation] Tools used for interpolations, mainly for log grids.
        \item [physical\_systems] Structures and necessary functions for the simulations of wave systems. We use a sub-folder for each system, which contains: 
        \begin{description}
            \item [basics.jl] Definitions of the dispersion relation, interaction coefficients, resonant manifold, etc.
            \item [collision\_integral.jl] Defines the method for computing for computing the collision integral $St_k$. The method is overloaded, allowing the same syntax for all solvers.
            \item [structure.jl] Main Julia structure for the physical system.
            \item [diagnostics.jl] Non-default diagnostics/functions.
        \end{description}
        Common definitions and structures used for several wave systems are given in \texttt{basics\_all.jl}. Some wave systems contain multiple variations grouped in the same sub-folder for consistency (e.g., \texttt{Acoustic} includes $2D$ and $3D$ isotropic acoustic wave systems with the same dispersion relation but differing in spatial dimension and/or forcing).
        \item [plot] Standard \texttt{WavKins} plots, created using the \href{https://docs.makie.org/stable/}{Makie.jl} package with the \texttt{GLMakie.jl} backend. 
        \item [time\_stepping] Standard functions for time stepping.
        \item [WavKinS\_diagnostics.jl] Definitions of default diagnostics (total waveaction, total energy, total energy dissipation, energy spectrum, waveaction fluxes, energy fluxes, etc.).
        \item [WavKinS\_outputs.jl] Input/Output functions.
        \item [WavKinS\_parameters.jl] Structures for simulation parameters (output directory, output selection, final time, etc.).
        \item [WavKinS\_partitions\_threads.jl] Tools for multithreading, including functions for partitioning nodes among threads.
        \item [WavKinS\_structures.jl] Basic structures used in \WavKinS simulations (spectra, parameter containers, etc.).
    \end{description}
    \item[run]  Run simulations and test the code.
    \begin{description}
        \item[simple]  Simple simulations for various physical systems. 
        \item[tests] Test the different parts of the code (see the folder \texttt{src}) and next section.
    \end{description}
    \item[postproc] Example scripts (Julia, Python, and Matlab) that show how we can analyze outputs of some simulations.
    \item[docs] Documentation.
\end{description}

% #https://excalidraw.com
% ```mermaid
% graph TD
%     A[Initialize Parameters] --> B[Create Grid]
%     B --> C[Define Physical System]
%     C --> D[Set Solver Parameters]
%     D --> E[Run Solver]
%     E --> F[Compute Diagnostics]
%     F --> G[Analyze Results]
% ```

\subsection{Quality control}

The current version of the code contains tests of the integration and interpolation schemes. Moreover, each physical system contains scripts to test the conservation of their respective invariants. The current version of the code does not have yet automatic unit tests.

\section*{Availability}
% \vspace{0.5cm}
 \subsection*{Operating system}\vspace{-0.25cm}
This solver can be run on any operating system that supports Julia (Linux, MacOS, Windows).

\subsection*{Programming language}\vspace{-0.25cm}
Julia 1.10.5 or higher

\subsection*{Additional system requirements}
\vspace{-0.25cm}
None
 \subsection*{Dependencies} \vspace{-0.25cm}
 Julia 1.10.5 or higher, and the following packages: BSplineKit v0.17.7, BenchmarkTools v1.5.0, CairoMakie v0.12.16, Documenter v1.8.0, GLMakie v0.10.16, LaTeXStrings v1.4.0, NCDatasets v0.14.6, PProf v3.1.3, TimerOutputs v0.5.25, Trapz v2.0.3.
%
 %
% \textcolor{blue}{E.g. libraries, frameworks, incl. minimum version compatibility.}

\subsection*{List of contributors}
\vspace{-0.25cm}
The current contributors of \WavKinS are Giorgio Krstulovic and Vincent Labarre.

\subsection*{Software location:}

\begin{description}[noitemsep,topsep=0pt]
    \item[Archive] ~
    \begin{description}[noitemsep,topsep=0pt]
        \item[Name:] Zenodo 
        \item[Persistent identifier:] \url{https://doi.org/10.5281/zenodo.15092688}
        \item[Licence:] EUPL-1.2 license agreement.
        \item[Publisher:]  Giorgio Krstulovic.
        \item[Version published:] 0.1.1
        \item[Date published:] 26/03/2025
    \end{description}
\end{description}

\vspace{.25cm}
\begin{description}[noitemsep,topsep=0pt]
    \item[Code repository] ~
    \begin{description}[noitemsep,topsep=0pt]
        \item[Name:] GitHub
        \item[Persistent identifier:] \url{https://github.com/gkrstulovic/WavKinS.jl}
        \item[Licence:] EUPL-1.2 license agreement.
        \item[Date published:] 09/12/2024
    \end{description}
\end{description}

\subsection*{Language} \vspace{-0.25cm}
English

\section*{Reuse potential}

This code is a tool for simulating the wave kinetic equations of several physical systems, and can be thus used in scientific research.
To this date, \WavKinS has been used in two scientific publications \cite{labarre2024wave,zhu2024turbulence}. Any user with some basic experience in Julia can use the script provided in \WavKinS to completely reproduce the numerical results of these two works.

Several other physical systems are already implemented in \WavKinS and ready to be used. \WavKinS has been designed in a modular manner such that a user could use the basic structures of the code to easily implement new physical systems. In the documentation, we provide some basic guidelines on how this can be done, and give some general advice and good practices. There is no support for \WavKinSns. However, users can contact the authors of this work for eventual help or new collaborations.

% \textcolor{blue}{Please describe in as much detail as possible the ways in which the software could be reused by other researchers both within and outside of your field. This should include the use cases for the software, and also details of how the software might be modified or extended (including how contributors should contact you) if appropriate. Also you must include details of what support mechanisms are in place for this software (even if there is no support).}

\section*{Acknowledgements}
We acknowledge Guillaume Costa, Kiran Kolloru, Ying Zhu for their contributions to the code and Juan I. Polanco for his infinite help with Julia.

\section*{Funding statement}

This work was funded by the Simons Foundation Collaboration grant Wave Turbulence (Award ID 651471). 

\section*{Competing interests}

The authors declare that they have no competing interests.

\section*{References}

\bibliographystyle{apalike}
\bibliography{refs}

\begin{thebibliography}{}

\bibitem[Dyachenko et~al., 1992]{dyachenko1992optical}
Dyachenko, S., Newell, A., Pushkarev, A., and Zakharov, V. (1992).
\newblock {Optical turbulence: weak turbulence, condensates and collapsing filaments in the nonlinear Schr{\"o}dinger equation}.
\newblock {\em Physica D: Nonlinear Phenomena}, 57(1-2):96--160.

\bibitem[Galtier, 2003]{Galtier_WeakInertialwaveTurbulence_2003}
Galtier, S. (2003).
\newblock Weak inertial-wave turbulence theory.
\newblock {\em Physical Review E}, 68(1):015301.

\bibitem[Griffin et~al., 2022]{griffinEnergySpectrumTwoDimensional2022}
Griffin, A., Krstulovic, G., L’vov, V.~S., and Nazarenko, S. (2022).
\newblock Energy {Spectrum} of {Two}-{Dimensional} {Acoustic} {Turbulence}.
\newblock {\em Physical Review Letters}, 128(22):224501.

\bibitem[Labarre et~al., 2024a]{labarre2024wave}
Labarre, V., Krstulovic, G., and Nazarenko, S. (2024a).
\newblock Wave-kinetic dynamics of forced-dissipated turbulent internal gravity waves.
\newblock {\em arXiv preprint arXiv:2407.11469}.

\bibitem[Labarre et~al., 2024b]{Labarre_Lanchon_Cortet_Krstulovic_Nazarenko_2024}
Labarre, V., Lanchon, N., Cortet, P.-P., Krstulovic, G., and Nazarenko, S. (2024b).
\newblock On the kinetics of internal gravity waves beyond the hydrostatic regime.
\newblock {\em Journal of Fluid Mechanics}, 998:A17.

\bibitem[L'vov and Nazarenko, 2010]{Lvov_WeakTurbulenceKelvin_2010}
L'vov, V.~S. and Nazarenko, S. (2010).
\newblock Weak turbulence of {{Kelvin}} waves in superfluid {{He}}.
\newblock {\em Low Temperature Physics}, 36(8):785--791.

\bibitem[Lvov et~al., 2012]{lvovResonantNearResonantInternal2012a}
Lvov, Y.~V., Polzin, K.~L., and Yokoyama, N. (2012).
\newblock Resonant and {Near}-{Resonant} {Internal} {Wave} {Interactions}.
\newblock {\em Journal of Physical Oceanography}, 42(5):669--691.

\bibitem[Lvov and Tabak, 2001]{lvovHamiltonianFormalismGarrettMunk2001}
Lvov, Y.~V. and Tabak, E.~G. (2001).
\newblock Hamiltonian {Formalism} and the {Garrett}-{Munk} {Spectrum} of {Internal} {Waves} in the {Ocean}.
\newblock {\em Physical Review Letters}, 87(16):168501.

\bibitem[Majda et~al., 1997]{majdaOnedimensionalModelDispersive1997}
Majda, A.~J., McLaughlin, D.~W., and Tabak, E.~G. (1997).
\newblock A one-dimensional model for dispersive wave turbulence.
\newblock {\em Journal of Nonlinear Science}, 7(1):9--44.

\bibitem[Nazarenko, 2011]{nazarenko2011wave}
Nazarenko, S. (2011).
\newblock {\em Wave turbulence}, volume 825.
\newblock Springer Science \& Business Media.

\bibitem[Zakharov et~al., 2004]{zakharovOnedimensionalWaveTurbulence2004}
Zakharov, V., Dias, F., and Pushkarev, A. (2004).
\newblock One-dimensional wave turbulence.
\newblock {\em Physics Reports}, 398(1):1--65.

\bibitem[Zakharov et~al., 1992]{ZLF}
Zakharov, V.~E., L'vov, V.~S., and Falkovich, G. (1992).
\newblock {\em {Kolmogorov spectra of turbulence 1: Wave turbulence}}.
\newblock Springer-Verlag.

\bibitem[Zhu et~al., 2024]{zhu2024turbulence}
Zhu, Y., Krstulovic, G., and Nazarenko, S. (2024).
\newblock Turbulence and far-from-equilibrium equation of state of bogoliubov waves in bose-einstein condensates.
\newblock {\em arXiv preprint arXiv:2408.15163}.

\bibitem[Zhu et~al., 2023]{zhuDirectInverseCascades2023}
Zhu, Y., Semisalov, B., Krstulovic, G., and Nazarenko, S. (2023).
\newblock Direct and {Inverse} {Cascades} in {Turbulent} {Bose}-{Einstein} {Condensates}.
\newblock {\em Physical Review Letters}, 130(13):133001.

\end{thebibliography}

\end{document}